\documentclass[journal]{IEEEtran}

\ifCLASSINFOpdf
\else
   \usepackage[dvips]{graphicx}
\fi
\usepackage{url}

\usepackage{graphicx}

\usepackage{amsmath}
\usepackage{amssymb}
\usepackage{color}

\ifCLASSOPTIONcompsoc
	\usepackage[caption=false,font=normalsize,labelfont=sf,textfont=sf]{subfig}
\else
	\usepackage[caption=false,font=footnotesize]{subfig}
\fi

\begin{document}

\title{Clutter Edges Detection Algorithms for Structured Clutter Covariance Matrices}

\author{Tianqi Wang, Da Xu, \IEEEmembership{Member, IEEE}, Chengpeng Hao, \IEEEmembership{Senior Member, IEEE}, Pia Addabbo, \IEEEmembership{Senior Member, IEEE}, and Danilo Orlando, \IEEEmembership{Senior Member, IEEE}
\thanks{This work was supported in part by the National Natural Science Foundation of China under Grant 61701489, 61971412, and 
by the IACAS Frontier Exploration Project under Grant QYTS202012.}
\thanks{Tianqi Wang is with the Institute of Acoustics, Chinese Academy of Sciences, Beijing 100190, China, and also with the School of Electronic, Electrical and Communication Engineering, University of Chinese Academy of Sciences, Beijing 100049, China (e-mail: wangtianqi@mail.ioa.ac.cn).}
\thanks{Da Xu and Chengpeng Hao are with the Institute of Acoustics, Chinese Academy of Sciences, Beijing 100190, China (e-mail: xuda@mail.ioa.ac.cn; haochengp@mail.ioa.ac.cn).}
\thanks{Pia Addabbo is with Universit\`{a} degli studi del Sannio, Piazza Roma, 21, Benevento 82100, Italy (e-mail: paddabbo@unisannio.it).}
\thanks{Danilo Orlando is with University ``Niccolò Cusano'', 00166 Rome, Italy (e-mail: danilor78@gmail.com).}
}

\markboth{Journal of \LaTeX\ Class Files, Vol. , No. , December 2021}
{Shell \MakeLowercase{\textit{et al.}}: Bare Demo of IEEEtran.cls for IEEE Journals}
\maketitle

\begin{abstract}
This letter deals with the problem of clutter edge detection and localization in training data. To this end, the problem is formulated as a binary hypothesis test assuming that the ranks of the clutter covariance matrix are known, and adaptive architectures are designed based on the generalized likelihood ratio test to decide whether the training data within a sliding window contains a homogeneous set or two heterogeneous subsets. In the design stage, we utilize four different covariance matrix structures (i.e., Hermitian, persymmetric, 
symmetric, and centrosymmetric) to exploit the a priori information. Then, for the case of unknown ranks, the architectures are extended by devising a preliminary estimation stage resorting to the model order selection rules. Numerical examples based on both synthetic and real data highlight that the proposed solutions possess superior detection and localization performance with respect to the competitors that do not use any a priori information.
\end{abstract}

\begin{IEEEkeywords}
Adaptive radar detection, classification, clutter edge, covariance structure, generalized likelihood ratio test.
\end{IEEEkeywords}

\IEEEpeerreviewmaketitle

\vspace{-5mm}

\section{Introduction}
\label{Sec:Introduction}

\IEEEPARstart{O}{ver} the last decades, adaptive radar detection of targets in Gaussian interference 
with unknown covariance matrix is an active field of research and many solutions have been proposed such as \cite{1}-\cite{19}. 
Most of the aforementioned solutions are based on the so-called Homogeneous Environment (HE) where a set of training samples,
used for estimation purposes and collected in the proximity of the Cell Under Test (CUT), is assumed to share the same spectral properties of the interference in the CUT. However, in practical applications, clutter background can be characterized by unknown clutter edges, 
namely the interference power between adjacent range cells varies and, hence, the statistical properties 
of the training samples do not meet the HE assumption seriously impacting on the performance of the traditional detection methods.

To solve the above problem, several contributions can be found in the related literature such as \cite{20}-\cite{23}.
In the most recent work \cite{23}, Xu et al. proposed a Known Rank-Clutter Edge Detector (KR-CED) to perform 
the adaptive detection and localization of the clutter edge also addressing the case of unknown rank 
of the Clutter Covariance Matrix (CCM). 
Therein, the so-called Covariance Change Detector (CCD) has been considered
as a competitor; such a decision scheme is a straightforward extension of the Generalized Likelihood Ratio Test (GLRT) for equality 
of covariance matrices \cite{24} to the complex domain and assuming that the location of the change point is unknown.

This letter enriches the framework of \cite{23} with unexplored (at least to the best of authors' knowledge) cases. 
Unlike \cite{23}, a more general scenario is considered where the ranks of the CCM in the regions on both sides of clutter edge can be different. 
In addition, different CCM structures are taken into account. 
Such structures arise from system geometries and/or clutter
properties \cite{25}-\cite{27}. 
In fact, a system using a symmetrically spaced linear array leads to a CCM that 
is Hermitian about its principal diagonal and persymmetric about its cross diagonal \cite{28}.
Another example is related to ground clutter observed by a stationary monostatic radar that gives rise to 
a symmetric power spectral density and, hence, a real-valued as well as even autocorrelation function (symmetric CCM) \cite{29}.
When both the above situations occur, then the CCM exhibits a centrosymmetric structure \cite{27}.
Thus, four CCM structures are considered: the generic unstructured (Hermitian) case; 
the persymmetric (or centrohermitian) case; the (real) symmetric case; and the (real) centrosymmetric case.   
The problem is formulated as a binary hypothesis test where the alternative hypothesis assumes the 
presence of a clutter edge. The design first assumes that the ranks of the CCMs are known.
Then, the proposed architectures are coupled with a preliminary stage for rank estimation based upon 
Model Order Selection (MOS) rules \cite{30}.
Finally, the performance analysis is conducted on both synthetic and recorded live data.

The remainder of this letter is organized as follows. Section \ref{Sec:ProblemFormulation} deals with the problem formulation while Section \ref{Sec:Detection} focuses on the designs. Section \ref{Sec:Numerical Examples} provides numerical examples and comparisons. 
Concluding remarks are given in Section \ref{Sec:Conclusion}.\footnote{Notation: vectors (matrices) are 
denoted by boldface lower- (upper-) case letters; $\det(\cdot)$, $\textrm{Tr}(\cdot)$, $\textrm{Rank}(\cdot)$, 
$(\cdot)^*$, $(\cdot)^T$, and $(\cdot)^\dag$ denote the determinant, the trace, the rank, complex conjugate, transpose, and complex conjugate transpose, respectively. $\mathbb{R}$ ($\mathbb{C}$) is the set of real (complex) numbers, $\mathbb{R}^{N\times M}$
($\mathbb{C}^{N\times M}$) is the Euclidean space of $(N\times M)$-dimensional real (complex) matrices (or vectors if $M=1$). 
$\mathfrak{Re}\{\cdot\}$ and $\mathfrak{Im}\{\cdot\}$ indicate the real and imaginary parts of a complex number, respectively. $\mathbf{I}$ and $\mathbf{0}$ stand for the identity matrix and the null matrix, respectively, of suitable sizes. $\mathbf{J}\in\mathbb{R}^{N\times N}$ denotes a permutation matrix such that $\mathbf{J}(l,k)=1$ only if $l+k=N+1$. $\mathbf{x} \sim \mathcal{CN}_N(\mathbf{\mu}, \mathbf{R})$ means that $\mathbf{x}$ is a complex circular $N$-dimensional normal vector with mean $\mathbf{\mu}$ and covariance matrix $\mathbf{R}$. For any Hermitian matrix $\mathbf{A}$, $\mathbf{A}\succeq\mathbf{0}$ means that $\mathbf{A}$ is a positive semi-definite matrix.}

\section{Problem Formulation}
\label{Sec:ProblemFormulation}

\begin{figure}[htbp]
	\centering
		\includegraphics[width=0.75\columnwidth,height=0.33\columnwidth]{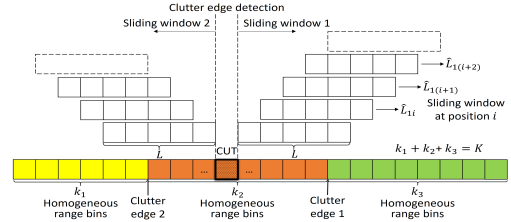}
		\caption{Two sliding windows of the same size $L$: the first moves forward and the second one moves backward from the CUT.}
		\label{fig_slidingWindows}
\end{figure}

The problem at hand consists in deciding for the presence of clutter edges in the area surrounding the CUT. 
To this end, we consider a radar system, equipped with $N\geq 2$ space, time, or space-time channels which 
illuminates the CUT and the surrounding area consisting of $K$ range bins. 
Let us assume that the set can contain clutter edges and consider two sliding windows 
of the same size $L<K/2$, 
which\footnote{
The choice of $L$ depends on several factors that range from computational requirements
to estimation quality. In fact, high values for $L$ increase the computational load but, at the same time, can
allow for high-quality estimation and vice versa. However, the drawback related to large values of $L$ might be the presence
of multiple clutter regions and, hence, multiple change points within the same sliding window.
} 
moves towards the two opposite directions (forward and backward) from the CUT as illustrated in Fig.
\ref{fig_slidingWindows}. Moving over the entire set of range bins, the radar system decides 
whether or not a clutter edge is present for both the sliding directions. 
The two clutter edge search operations can be conducted independently each from the other. 
Therefore, denoting by $\mathbf{z}_l \in \mathbb{C}^{N\times 1}, l=1,2,\cdots, L$, statistically independent 
data vectors belonging to one of the two sliding windows, the decision problem can be formulated as the following hypothesis testing problem 
\begin{equation}
	\label{eqn:detProb}
	\begin{cases}
		H_0 :
		~~\mathbf{z}_{l}\sim\mathcal{CN}_N(\mathbf{0},\sigma_n^2\mathbf{I}+\mathbf{M}_0),~ l=1,\cdots,L,\\
		H_1 :
		\begin{cases}
			\mathbf{z}_{l}\sim\mathcal{CN}_N(\mathbf{0},\sigma_n^2\mathbf{I}+\mathbf{M}_1),~ l=1,\cdots,L_1,\\			\mathbf{z}_{l}\sim\mathcal{CN}_N(\mathbf{0},\sigma_n^2\mathbf{I}+\mathbf{M}_2),~ l=L_1+1,\cdots,L,
		\end{cases}
	\end{cases}
\end{equation}
where $\sigma_n^2>0$ is the thermal noise power, $\mathbf{M}_0 \in \mathbb{C}^{N\times N}$ is the whole CCM under $H_0$ with $\textrm{Rank}(\mathbf{M}_0)=r_0<N$, $\mathbf{M}_1 \in \mathbb{C}^{N\times N}$  is the first region CCM component with $\textrm{Rank}(\mathbf{M}_1)=r_1<N$, $\mathbf{M}_2 \in \mathbb{C}^{N\times N}$ is the second region CCM component with $\textrm{Rank}(\mathbf{M}_2)=r_2<N$ and such that $\mathbf{M}_2-\mathbf{M}_1 \succeq \mathbf{0}$, or $\mathbf{M}_1-\mathbf{M}_2 \succeq \mathbf{0}$, $L\geq r_1+r_2$ and $L_1 \in \left\{l_{1\textrm{min}},\dots,l_{1\textrm{max}}\right\}$ with $l_{1\textrm{min}} = \max\{r_1,r_2\}+1$, $l_{1\textrm{max}} = L-\max\{r_1,r_2\}-1$.

Then the probability density function of $\mathbf{Z}=[\mathbf{z}_1,\ldots,\mathbf{z}_L]\in\mathbb{C}^{N\times L}$ under $H_0$ has the following expression
\begin{equation}
	\label{eqn:pdf_H0}
	f(\mathbf{Z};\sigma_{n}^2,\mathbf{M}_0,H_0)=
	\frac{\exp{\left\{-\textrm{Tr}[(\sigma_{n}^2 \mathbf{I}+\mathbf{M}_0)^{-1}\mathbf{S}_0]\right\}}}
	{\left[\pi^{N}\det(\sigma_{n}^2\mathbf{I}+ \mathbf{M}_0)\right]^{L}},
\end{equation}
where $\mathbf{S}_0 = \mathbf{Z}\mathbf{Z}^\dag$, while that under $H_1$ is given by 
\begin{align}
	\label{eqn:pdf_H1}
	\nonumber
	f(\mathbf{Z};\sigma_n^2,\mathbf{M}_1,\mathbf{M}_2,L_1&,H_1)=
	\frac{\exp{\left\{-\textrm{Tr}[(\sigma_{n}^2 \mathbf{I}+\mathbf{M}_1)^{-1}\mathbf{S}_1]\right\}}}
	{\left[\pi^{N}\det(\sigma_{n}^2\mathbf{I}+\mathbf{M}_1)\right]^{L_1}}
	\\
	&\times
	\frac{\exp{\left\{-\textrm{Tr}[(\sigma_{n}^2 \mathbf{I}+\mathbf{M}_2)^{-1}\mathbf{S}_2]\right\}}}
	{\left[\pi^{N}\det(\sigma_{n}^2\mathbf{I}+\mathbf{M}_2)\right]^{L_2}},
\end{align}
where $L_2=L-L_1$, $\mathbf{S}_1 \negthinspace=\negthinspace \sum_{l=1}^{L_1}\mathbf{z}_l\mathbf{z}_l^\dag$, and $\mathbf{S}_2 \negthinspace=\negthinspace \sum_{l=L_1+1}^{L}\mathbf{z}_l\mathbf{z}_l^\dag$. 

As aforementioned, we assume that $\mathbf{M}_i, i=0,1,2$, can be either Hermitian, persymmetric such that $\mathbf{M}_i^{-1}=[\mathbf{M}_i^{-1}+\mathbf{J}(\mathbf{M}_i^*)^{-1}\mathbf{J}]/2$, real symmetric implying that $\mathfrak{Re}\{\mathbf{z}_l\}$ and $\mathfrak{Im}\{\mathbf{z}_l\}$ are Independent Identically Distributed (IID), or centrosymmetric such that $\mathbf{M}_i^{-1}=[\mathbf{M}_i^{-1}+\mathbf{J}\mathbf{M}_i^{-1}\mathbf{J}]/2\in\mathbb{R}^{N\times N}$ implying that $\mathfrak{Re}\{\mathbf{z}_l\}$ and $\mathfrak{Im}\{\mathbf{z}_l\}$ are IID. Considering $\mathbf{S}_i, i=0,1,2$, defined above, we can write the following equalities:
\begin{align}
	\label{eqn:ICMsprop}
	& \textrm{Tr}\left[ \left( \sigma_n^2 \mathbf{I}+\mathbf{M}_i\right)^{-1} \mathbf{S}_i  \right]  \\ \nonumber
	&= \begin{cases}
		\textrm{Tr}\left[ \left( \sigma_n^2 \mathbf{I}+\mathbf{M}_i\right)^{-1} \left( \mathbf{S}_i  +\mathbf{J}\mathbf{S}_i^*\mathbf{J} \right)/2 \right],
		\\ \text{ if $\mathbf{M}_i$ is persymmetric};
		\\
		\textrm{Tr}\left[ \left( \sigma_n^2 \mathbf{I}+\mathbf{M}_i\right)^{-1}  \mathfrak{Re}\{\mathbf{S}_i\} \right], 
		\text{ if $\mathbf{M}_i$ is real symmetric};
		\\
		\textrm{Tr}\left[ \left( \sigma_n^2 \mathbf{I}+\mathbf{M}_i\right)^{-1} \left( \mathfrak{Re}\{\mathbf{S}_i\}  +\mathbf{J}\mathfrak{Re}\{\mathbf{S}_i^*\} \mathbf{J} \right)/2 \right],
		\\ \text{ if $\mathbf{M}_i$ is centrosymmetric}.
	\end{cases}
\end{align}

\section{Detection Architecture Designs}
\label{Sec:Detection}

In this section, assuming that a specific position of the sliding window is given, we design detection architectures that can declare the presence of a clutter edge and provide an estimate of its position within the considered window. We start the design under the assumption of known $\mathbf{r}=\left[r_0,r_1,r_2\right]^T$, then we deal with the case of unknown $\mathbf{r}$.

\subsection{Design for known $\textbf{r}$}
\label{Subsec:known_r}

In order to solve problem \eqref{eqn:detProb} assuming that $\mathbf{r}$ is known, we utilize the GLRT, whose expression is
{
\begin{equation}
	\Lambda_{\mathbf{r}}(\mathbf{Z})=
	\frac{\max\limits_{L_1}\max\limits_{\sigma^2_n}\max\limits_{\mathbf{M}_1}\max\limits_{\mathbf{M}_2}f(\mathbf{Z};\sigma_n^2,\mathbf{M}_1,\mathbf{M}_2,L_1,H_1)}
	{\max\limits_{\sigma^2_n}\max\limits_{\mathbf{M}_0}f(\mathbf{Z};\sigma_{n}^2,\mathbf{M}_0,H_0)}
	\overset{H_1}{\underset{H_0}{\gtrless}}\eta,
\end{equation}
}
where $\eta$ is the detection threshold.\footnote{Hereafter, we denote by $\eta$ the generic detection threshold set according to the desired Probability of False Edge Detection ($P_{\textrm{FED}}$).}

Under $H_0$, the log-likelihood of $\mathbf{Z}$ can be expressed as
\begin{multline}
	\log f(\mathbf{Z};\sigma_{n}^2,\mathbf{M}_0,H_0)
	= C-L\log(\det(\sigma_{n}^2\mathbf{I}+ \mathbf{M}_0)) 
	\\
	-\textrm{Tr}\left[ \left( \sigma_n^2 \mathbf{I}+\mathbf{M}_0\right)^{-1} \mathbf{S}_{h0} \right],
\end{multline}
where $C=-LN\log\pi$ and
\begin{equation*}
	\label{eqn:ICMsprop2}
	\mathbf{S}_{h0}\!=\!  
	\begin{cases}
		\mathbf{S}_0, \!\text{ for $\mathbf{M}_0$ hermitian};\\
		\left(\mathbf{S}_0 +\mathbf{J}\mathbf{S}_0^*\mathbf{J} \right)/2, \!\text{ for $\mathbf{M}_0$ persymmetric};\\
		\mathfrak{Re}\{\mathbf{S}_0\}, \!\text{ for $\mathbf{M}_0$ real symmetric};\\
		\left(\mathfrak{Re}\{\mathbf{S}_0\}  +\mathbf{J}\mathfrak{Re}\{\mathbf{S}_0^*\} \mathbf{J} \right)/2, \!\text{ for $\mathbf{M}_0$ centrosymmetric}.
	\end{cases}
\end{equation*}

The maximization of $\log f(\mathbf{Z};\sigma_{n}^2,\mathbf{M}_0,H_0)$ with respect to $\sigma_{n}^2$ and $\mathbf{M}_0$ can be accomplished following the lead of \cite{23} and \cite{31} to come up with 
\begin{equation}
	\label{eqn:l0}
	\hat{l}_0(r_0) \!= C \!-\! LN \!-\!L\left\{\sum\nolimits_{i=1}^{r_0}\log\frac{\gamma_i}{L}\!+\!(N \!-\! r_0)\log\widehat{\sigma}_{n,0}^2\right\},
\end{equation}
where $\gamma_1 \geq \gamma_2 \geq \ldots \geq \gamma_N \geq 0$ are the eigenvalues of $\mathbf{S}_{h0}$, and
$\widehat{\sigma}_{n,0}^2 = \frac{1}{L(N-r_0)} \sum\nolimits_{i=r_0+1}^{N}\gamma_i$.

Under $H_1$, the log-likelihood of $\mathbf{Z}$ can be expressed as
{ 
\begin{align}
	\nonumber
	\log f(\mathbf{Z};\sigma_n^2,\mathbf{M}_1,\mathbf{M}_2,L_1,H_1)
	&= C - L_1\! \log(\det(\sigma_{n}^2\mathbf{I}+ \mathbf{M}_1))
	\\ \nonumber
	-L_2\log(\det(\sigma_{n}^2\mathbf{I}+ \mathbf{M}_2)) 
	&- \textrm{Tr}\left[ \left( \sigma_n^2 \mathbf{I}+\mathbf{M}_1\right)^{-1} \mathbf{S}_{h1} \right]
	\\
	- &\textrm{Tr}\left[ \left( \sigma_n^2 \mathbf{I}+\mathbf{M}_2\right)^{-1} \mathbf{S}_{h2} \right],
\end{align}
}
where $\mathbf{S}_{hi}, i=1,2$ is defined as
\begin{equation*}
	\label{eqn:ICMsprop3}
	\mathbf{S}_{hi}\!=\!  
	\begin{cases}
		\mathbf{S}_i, \!\text{ for $\mathbf{M}_i$ hermitian};\\
		\left(\mathbf{S}_i +\mathbf{J}\mathbf{S}_i^*\mathbf{J} \right)/2, \!\text{ for $\mathbf{M}_i$ persymmetric};\\
		\mathfrak{Re}\{\mathbf{S}_i\}, \!\text{ for $\mathbf{M}_i$ real symmetric};\\
		\left(\mathfrak{Re}\{\mathbf{S}_i\}  +\mathbf{J}\mathfrak{Re}\{\mathbf{S}_i^*\} \mathbf{J} \right)/2, \!\text{ for $\mathbf{M}_i$ centrosymmetric}.
	\end{cases}
\end{equation*}
As done under $H_0$, maximizing $\log f(\mathbf{Z};\sigma_n^2,\mathbf{M}_1,\mathbf{M}_2,L_1,H_1)$
with respect to $\sigma_{n}^2$, $\mathbf{M}_1$ and $\mathbf{M}_2$ leads to
\begin{multline}
	\label{eqn:l1}
	\hat{l}_1(r_1,r_2,L_1) = C - LN -  \left[ L_1(N-r_1)+L_2(N-r_2)\right] 
	\\
	 \times \log\widehat{\sigma}_{n,1}^2 
	 - L_1\sum\nolimits_{i=1}^{r_1}\log{\frac{\gamma_{1i}}{L_1}} -L_2\sum\nolimits_{i=1}^{r_2}\log{\frac{\gamma_{2i}}{L_2}},
\end{multline}
where $\gamma_{11} \geq \gamma_{12} \geq \ldots \geq \gamma_{1N} \geq 0$ are the eigenvalues of $\mathbf{S}_{h1}$, $\gamma_{21} \geq \gamma_{22} \geq \ldots \geq \gamma_{2N} \geq 0$ are the eigenvalues of $\mathbf{S}_{h2}$ and 
$\widehat{\sigma}_{n,1}^2 = \frac{1}{L_1(N-r_1)+L_2(N-r_2)} \left( \sum\nolimits_{i=r_1+1}^{N}\gamma_{1i} +\sum\nolimits_{i=r_2+1}^{N}\gamma_{2i} \right)$.

Finally, gathering all the above derivations, we can write the logarithm of the GLRT for known $\mathbf{r}$ as
\begin{equation}
	\label{eqn:test}
	\log\Lambda_{\mathbf{r}}(\mathbf{Z}) \overset{H_1}{\underset{H_0}{\gtrless}}\eta,
\end{equation}
where 
$\log\Lambda_{\mathbf{r}}(\mathbf{Z}) =
\max\limits_{L_1\in\left\{l_{1\textrm{min}},\dots,l_{1\textrm{max}}\right\}} \left\{ \hat{l}_1(r_1,r_2,L_1) - \hat{l}_0(r_0) \right\} $
if ${\gamma_{1i}}/{L_1}>\widehat{\sigma}_{n,1}^2$, $\forall{i=1,\ldots,r_1}$ and ${\gamma_{2i}}/{L_2}>\widehat{\sigma}_{n,1}^2$, $\forall{i=1,\ldots,r_2}$, otherwise we set $\log\Lambda_{\mathbf{r}}(\mathbf{Z})=0$.

In what follows, we refer to \eqref{eqn:test} as Hermitian-Clutter Edge Detector (H-CED) when the CCMs are assumed general Hermitian, Persymmetric-Clutter Edge Detector (P-CED) when the CCMs have a persymmetric structure, Symmetric-Clutter Edge Detector (S-CED) when the clutter spectrum is symmetric with respect to the origin, and Centrosymmetric-Clutter Edge Detector (C-CED) when both spectrum symmetry and a persymmetric covariance structure hold.

In Fig. \ref{fig_slidingWindows}, when $H_1$ is declared for the $i$th position of the sliding window, an estimate of the clutter edge position, $\widehat{L}_{1i}$ say, within the sliding window is provided. As the sliding window moves, we will get several estimates. By using the fusion strategy in \cite{23}, the estimation quality can be improved.

\subsection{Implementation issues: unknown $\textbf{r}$}
\label{Subsec:r_estimate}

In this subsection, we provide a procedure to estimate 
the rank of $\mathbf{M}_i, i=0,1,2$ assuming the general Hermitian structure since it comprises the other considered structures. To this end, we resort to the MOS rules such as the Akaike Information Criterion (AIC), the Bayesian Information Criterion (BIC), and the Generalized Information Criterion (GIC), to come up with a suitable estimate for $\mathbf{r}$.

Precisely, under $H_0$, it is possible to estimate $r_0$ as $\widehat{r}_0=\arg\min\limits_{r_0}\left[-2\hat{l}_0(r_0)+q\cdot p(r_0)\right]$, where $p(r_0)=r_0(2N-r_0)+1$ is the number of unknown parameters \cite{32}, and
\begin{equation}
	q =
	\begin{cases}
		2~&\mbox{AIC penalty term},\\
		\log(L)~&\mbox{BIC-like penalty term},\\
		(1+a),~a>1~&\mbox{GIC penalty term}.\\
	\end{cases}
\end{equation}
Under $H_1$, for each allowable value of $L_1$, an estimate of $\mathbf{r}^{-}=[r_1,r_2]^T$ is obtained by computing
$\widehat{\mathbf{r}}^-_{L_1}=\arg\min\limits_{\mathbf{r}^-}\left[-2\hat{l}_1(r_1,r_2,L_1)+q\cdot\zeta(\mathbf{r}^-)\right]$, 
where $\zeta(\mathbf{r}^-)= 1 + \sum_{i=1}^{2} r_i (2N-r_i)$. Now, plugging the above estimates 
along with $\widehat{r}_0$ into $\log\Lambda_{\mathbf{r}}(\mathbf{Z})$,
where the last maximization is with respect to $L_1$, the final estimates of $L_1$ and $\mathbf{r}^-$ can be determined.

\section{Numerical Examples}
\label{Sec:Numerical Examples}

In this section, we present some numerical examples on simulated and real data. The considered performance metrics are the Probability of Edge Detection ($P_{\textrm{ED}}$) and Root Mean Square (RMS) estimation error of the clutter edge position.
The ranks are assumed known due to the excellent performance (not shown here for brevity) 
of the estimation procedure for the CPR values of interest.
A natural competitor of the proposed architectures is the CCD, whose expression is
\begin{equation}
	\label{eqn:competitor}
	\max\limits_{L_1\in\left\{l_{\rm{min}},\dots,l_{\rm{max}}\right\}}
	\frac{{[\det(\mathbf{S}_0/L)]}^L}
	{{[\det(\mathbf{S}_1/L_1)]}^{L_1}{[\det(\mathbf{S}_2/L_2)]}^{L_2}}
	\overset{H_1}{\underset{H_0}{\gtrless}}\eta,
\end{equation}
where $l_{\textrm{min}}>N$ and $L-l_{\textrm{max}}>N$. For $i=0,1,2$, if we consider the four cases of $\mathbf{M}_i$ and replace $\mathbf{S}_i$ with $\mathbf{S}_{hi}$, we will get Hermitian-CCD (H-CCD), Persymmetric-CCD (P-CCD), Symmetric-CCD (S-CCD) and Centrosymmetric-CCD (C-CCD). Moreover, the H-CED is also a competitor since it degenerates to the KR-CED derived in \cite{23} when $r_0\negthinspace=\negthinspace r_1\negthinspace=\negthinspace r_2$.

\subsection{Synthetic data}
\label{Subsec:Synthetic data}

In this part, we use standard Monte Carlo counting techniques by evaluating the detection thresholds and $P_{\textrm{ED}}$ (along with the RMS estimation errors) over 100/$P_{\textrm{FED}}$ and $10^4$ independent trials, respectively. The CCM is generated as 
$\mathbf{M}_i=\sigma_{c,i}^2\sum_{\theta_k\in\Theta_i}\mathbf{v}(\theta_k)\mathbf{v}(\theta_k)^\dag, i=0,1,2$, where $\sigma_{c,0}^2=\sigma_{c,1}^2$ is set based on the Clutter-to-Noise Ratio (CNR) defined as $\rm{CNR}=10\log_{10}{\left(\sigma_{c,0}^2/\sigma_n^2\right)}$ with $\sigma_n^2=1$, $\sigma_{c,2}^2$ is set based on the Clutter Power Ratio (CPR) defined as $\rm{CPR}=10\log_{10}{\left(\sigma_{c,2}^2/\sigma_{c,1}^2\right)}$, $\Theta_0=\Theta_1=\Theta_2=\{-20^{\circ},-10^{\circ},10^{\circ},20^{\circ}\}$ (implying that 
$r_{0}=r_{1}=r_{2}=4$ and $\mathbf{M}_i$ is centrosymmetric), and $\mathbf{v}(\theta_k)=
{\left[\textrm{exp}\{-j\pi\frac{N-1}{2}\sin{\theta_k}\},\dots,1,\dots,\textrm{exp}\{j\pi\frac{N-1}{2}\sin{\theta_k}\}\right]}^T\in\mathbb{C}^{N\times1}$ is the spatial steering vector. 
Moreover, we set $P_{\textrm{FED}}=10^{-4}$, $N=9$ and, for a fair comparison, the maximization over $L_1$ is carried out over the grid $\Omega\negthinspace=\negthinspace\{N+1,\dots,L-N-1\}$.

\begin{figure}[htbp]
	\centering
	\subfloat[$L_1=11$]{\includegraphics[width=0.49\columnwidth,height=0.33\columnwidth]{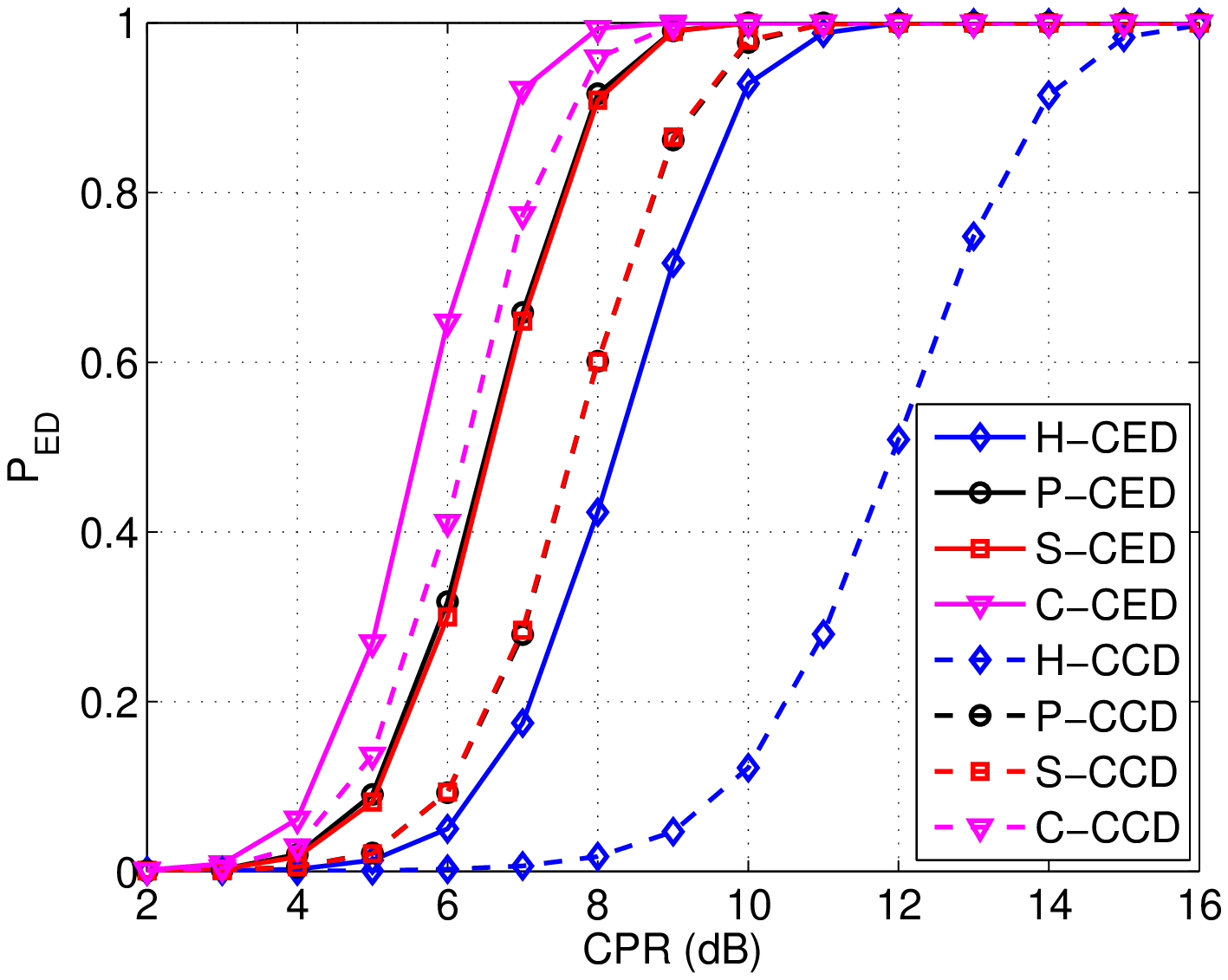}}
	\hfil
	\subfloat[$L_1=13$]{\includegraphics[width=0.49\columnwidth,height=0.33\columnwidth]{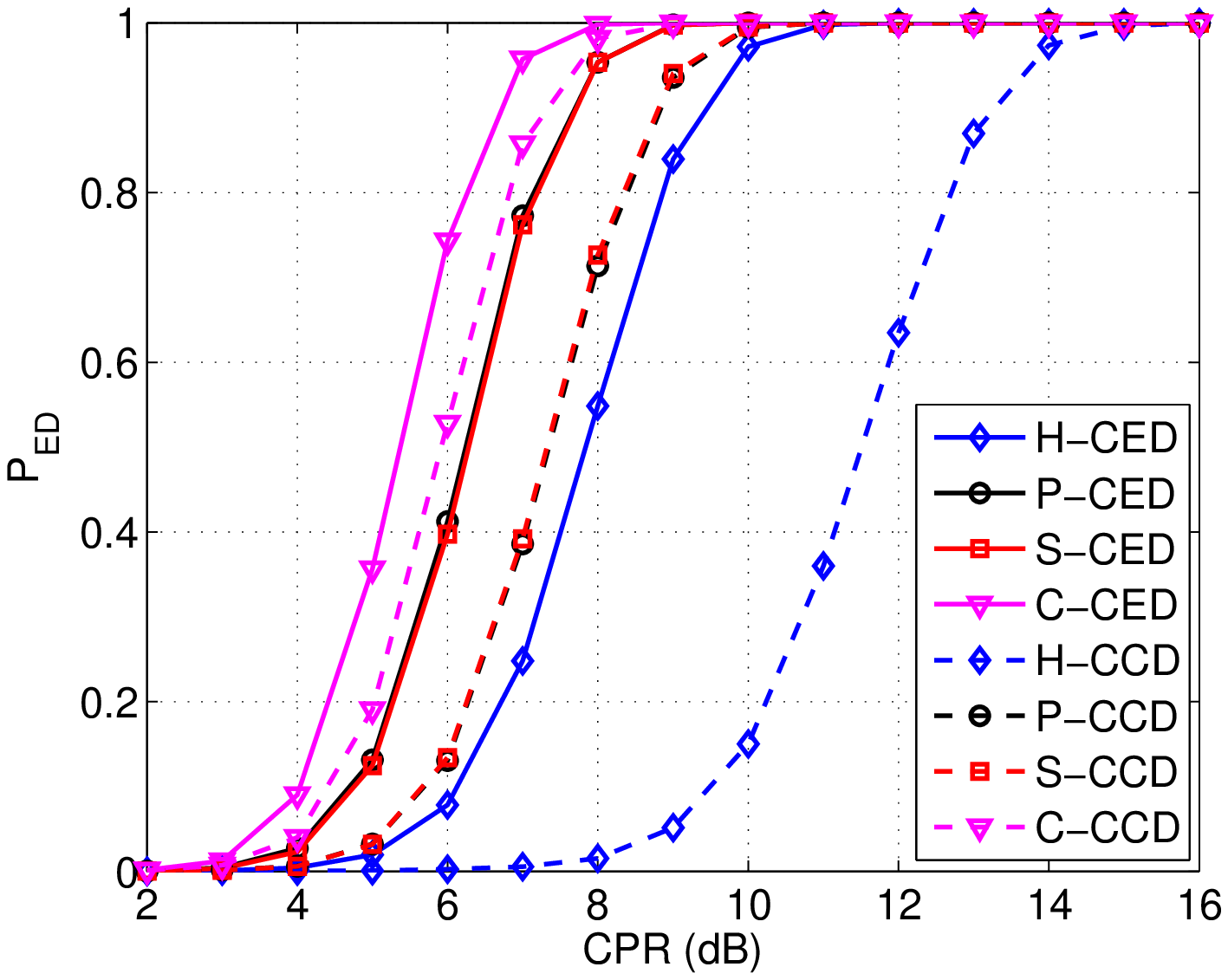}}
	\hfil
	\subfloat[$L_1=15$]{\includegraphics[width=0.49\columnwidth,height=0.33\columnwidth]{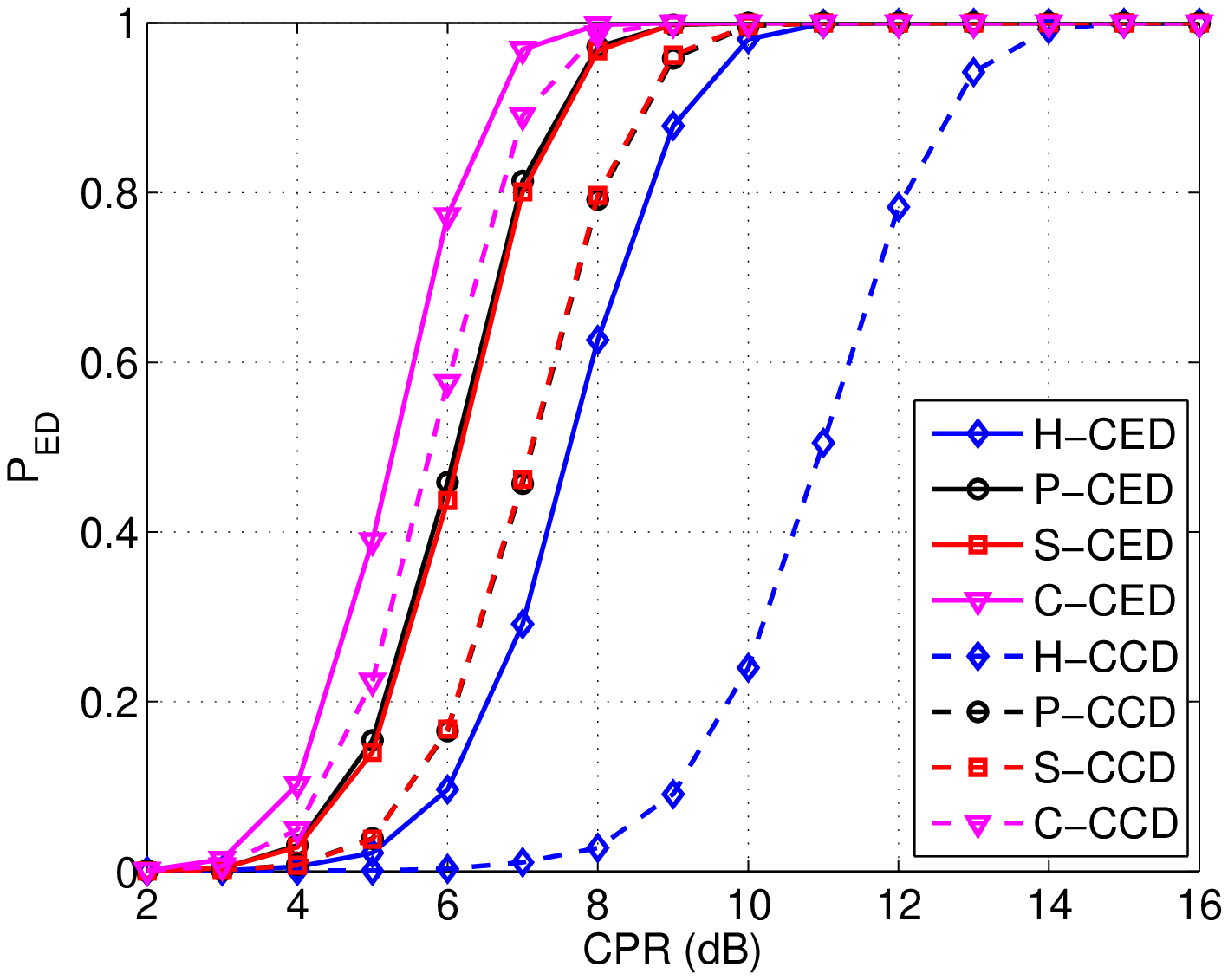}}
	\hfil
	\subfloat[$L_1=17$]{\includegraphics[width=0.49\columnwidth,height=0.33\columnwidth]{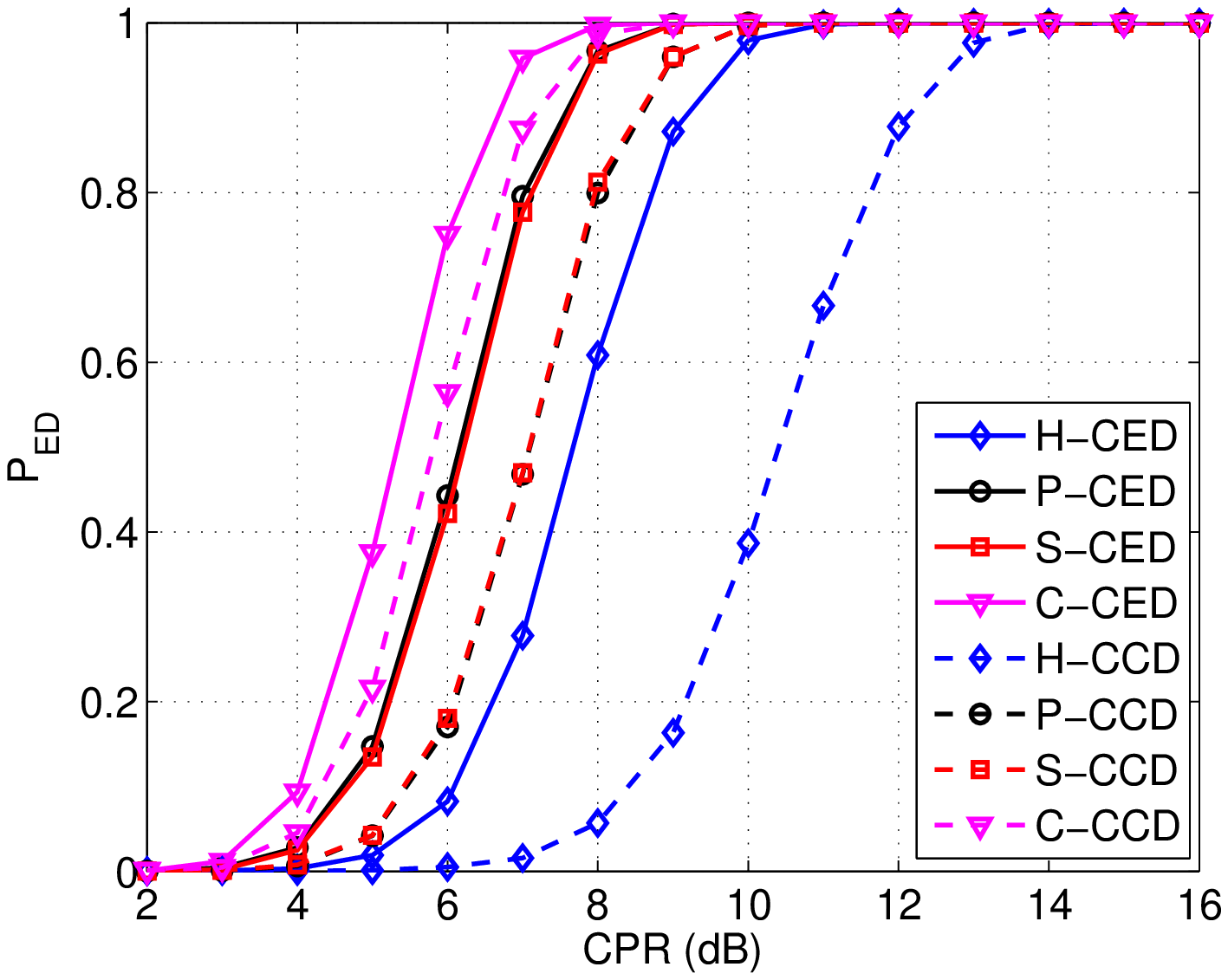}}
	\caption{$P_{\textrm{ED}}$ versus CPR for the proposed architectures and their competitors assuming $N=9$, $L=27$, $\rm{CNR}=25\;\rm{dB}$, and four values for $L_1$.}
	\label{fig_Ped}
\end{figure}

The clutter edge detection performance are shown in Fig. \ref{fig_Ped}, where C-CED has the best performance, followed by P-CED (along with S-CED) and H-CED, but all of them perform better than their corresponding competitors. Besides, the gains of the proposed architectures with respect to the competitors depend on the value of $L_1$ and attain their maximum when $L_1<L/2$. The algorithms are also capable to detect a change of the covariance structure. When the structures of $\mathbf{M}_1$ and $\mathbf{M}_2$ are different, $P_{\textrm{ED}}$ is 1 for all the CPR of interest (results are not reported here for brevity).

\begin{figure}[htbp]
	\centering
	\subfloat[$L=27$]{\includegraphics[width=0.49\columnwidth,height=0.33\columnwidth]{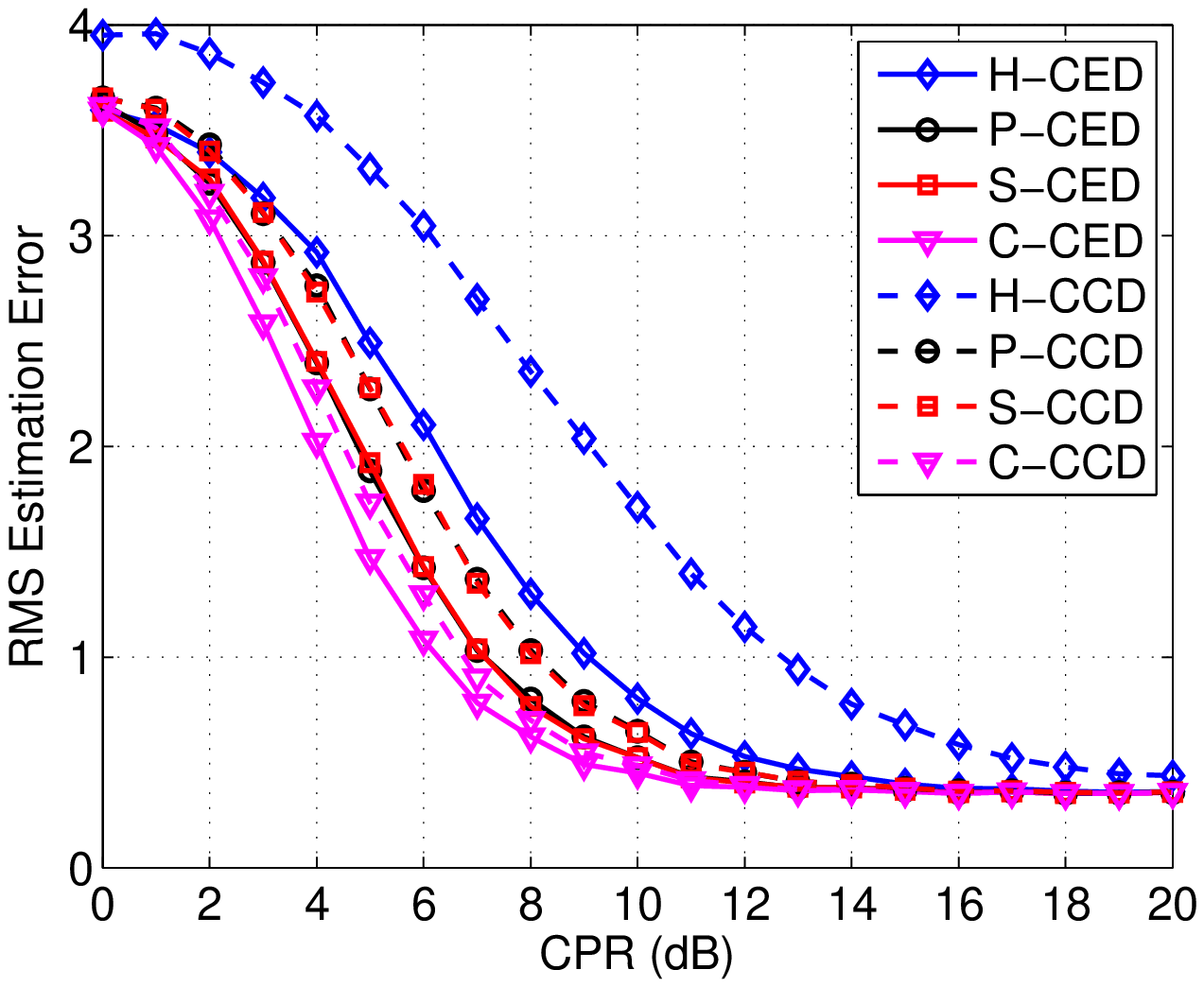}}
	\hfil
	\subfloat[$L=36$]{\includegraphics[width=0.49\columnwidth,height=0.33\columnwidth]{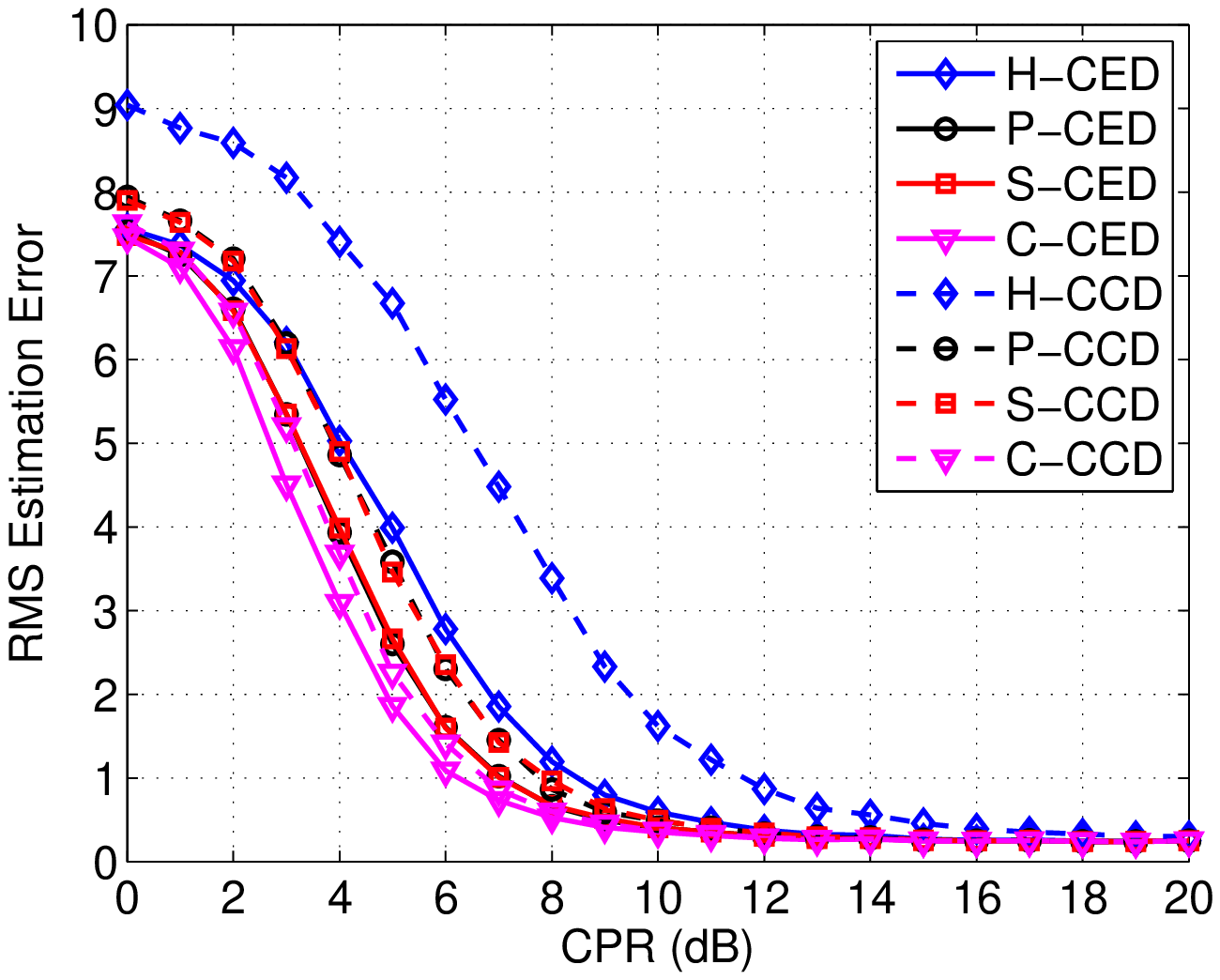}}
	\caption{RMS estimation error versus CPR for the proposed architectures and their competitors assuming $N=9$, $\rm{CNR}=25\;\rm{dB}$, and two values for $L$.}
	\label{fig_RMSE}
\end{figure}

In Fig. \ref{fig_RMSE}, we study the edge localization performance, where $L_1$ is generated as a discrete uniform random variable taking on values in $\Omega$. The plots show that for $\rm{CPR}<10\;\rm{dB}$, the errors of the proposed architectures are lower than that of their corresponding competitors. In addition, for low CPR, the error are larger for the sliding window with greater size.

Finally, notice that a performance degradation can occur when the actual CCM structure is a general case
of that assumed at the design stage. In fact, in this case, it is not ensured
that the actual CCM experiences the nominal structure.

\subsection{Real data}
\label{Subsec:Real data}

In this part, illustrative examples are based on the MIT-LL Phase-One radar dataset \cite{33},\cite{34}. 
Each acquisition is composed of 30720 temporal returns from 76 range bins. 
The covariance matrix of this dataset appears to have a centrosymmetric structure as 
noticed in \cite[and references therein]{35}.

The average power variation over the range (see Fig. 8 of [23]) indicates that there exist two almost uniform regions, occupying bins 1-10 and 22-37, separated by a transition region located around the range bin 12. In the following, we set $P_{\textrm{FED}}=10^{-3}$, $N=6$, and use $\widehat{\mathbf{r}}=[2,2,2]^T$ which has been already estimated in \cite{23} accounting also for numerical issues associated to the initial range bins and related to the system sensitivity time control. The detection threshold is computed based on the range bins 1-8 over 2560 data blocks.

\begin{figure}[htbp]
	\centering
	\subfloat[HH polarization]{\includegraphics[width=0.5\columnwidth,height=0.35\columnwidth]{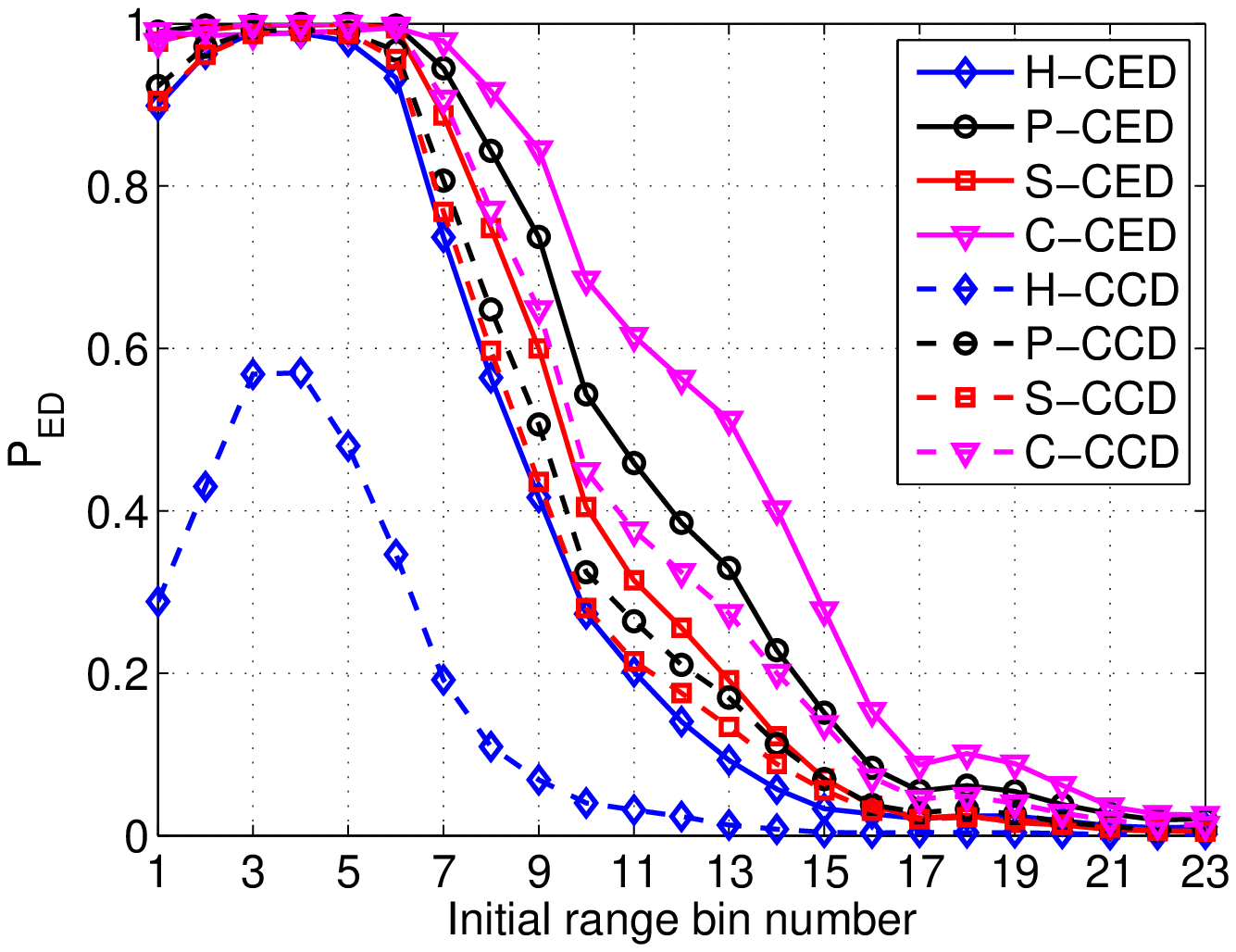}}
	\hfil
	\subfloat[VV polarization]{\includegraphics[width=0.5\columnwidth,height=0.35\columnwidth]{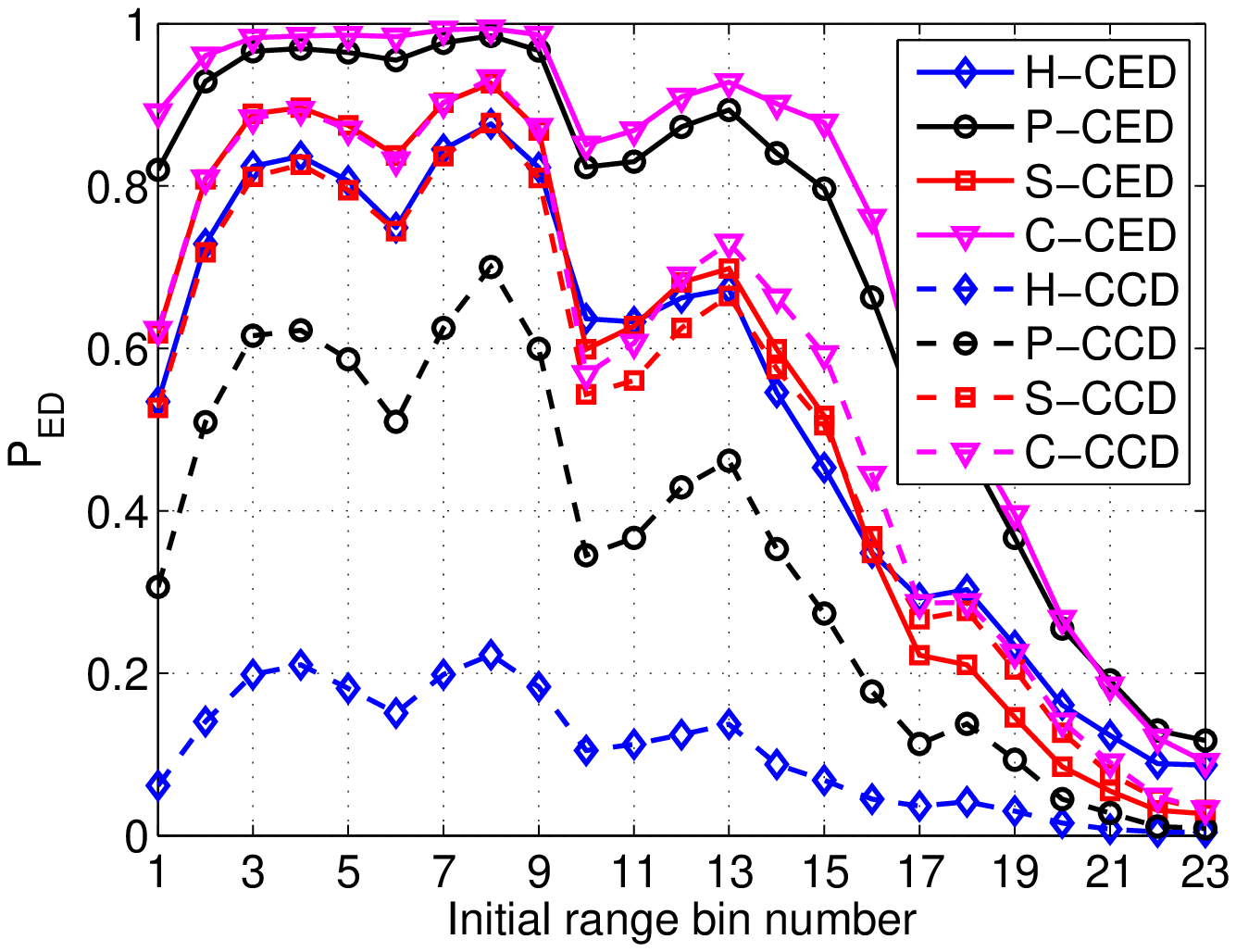}}
	\caption{$P_{\textrm{ED}}$ versus the initial range bin index of the sliding window for the proposed architectures and their competitors assuming $N=6$ and $L=16$.}
	\label{fig_PedRealData}
\end{figure}

Fig. \ref{fig_PedRealData} shows the detection performance based on 5120 data blocks, where the x-axis reports the initial bin index of the moving window. The plots show that the curves of the proposed architectures are above those of the competitors. All curves decrease when the window moves towards the transition region, which is less evident for data from the VV channel.

\begin{figure}[htbp]
	\centering
	\subfloat[HH polarization]{\includegraphics[width=0.5\columnwidth,height=0.35\columnwidth]{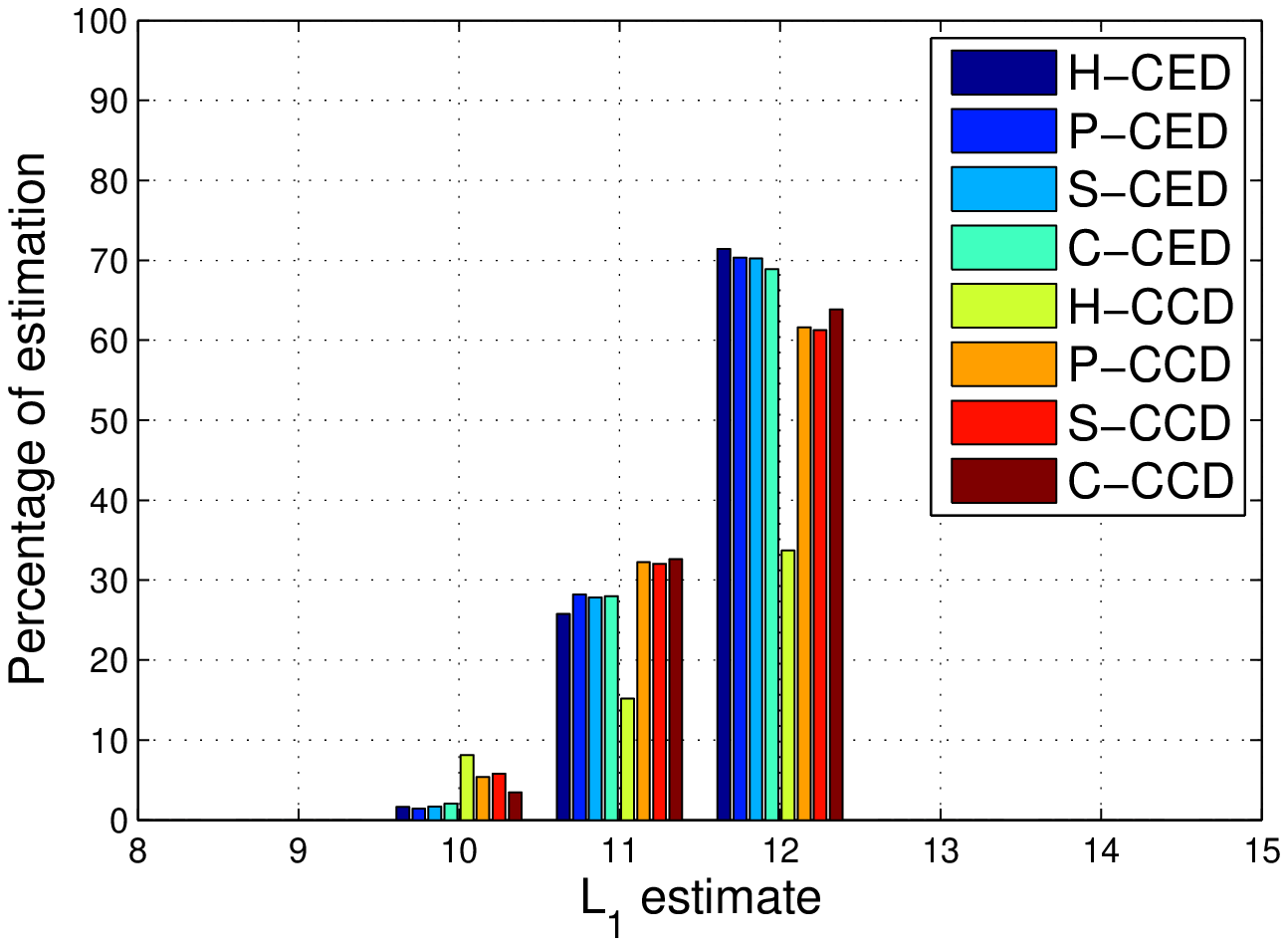}}
	\hfil
	\subfloat[VV polarization]{\includegraphics[width=0.5\columnwidth,height=0.35\columnwidth]{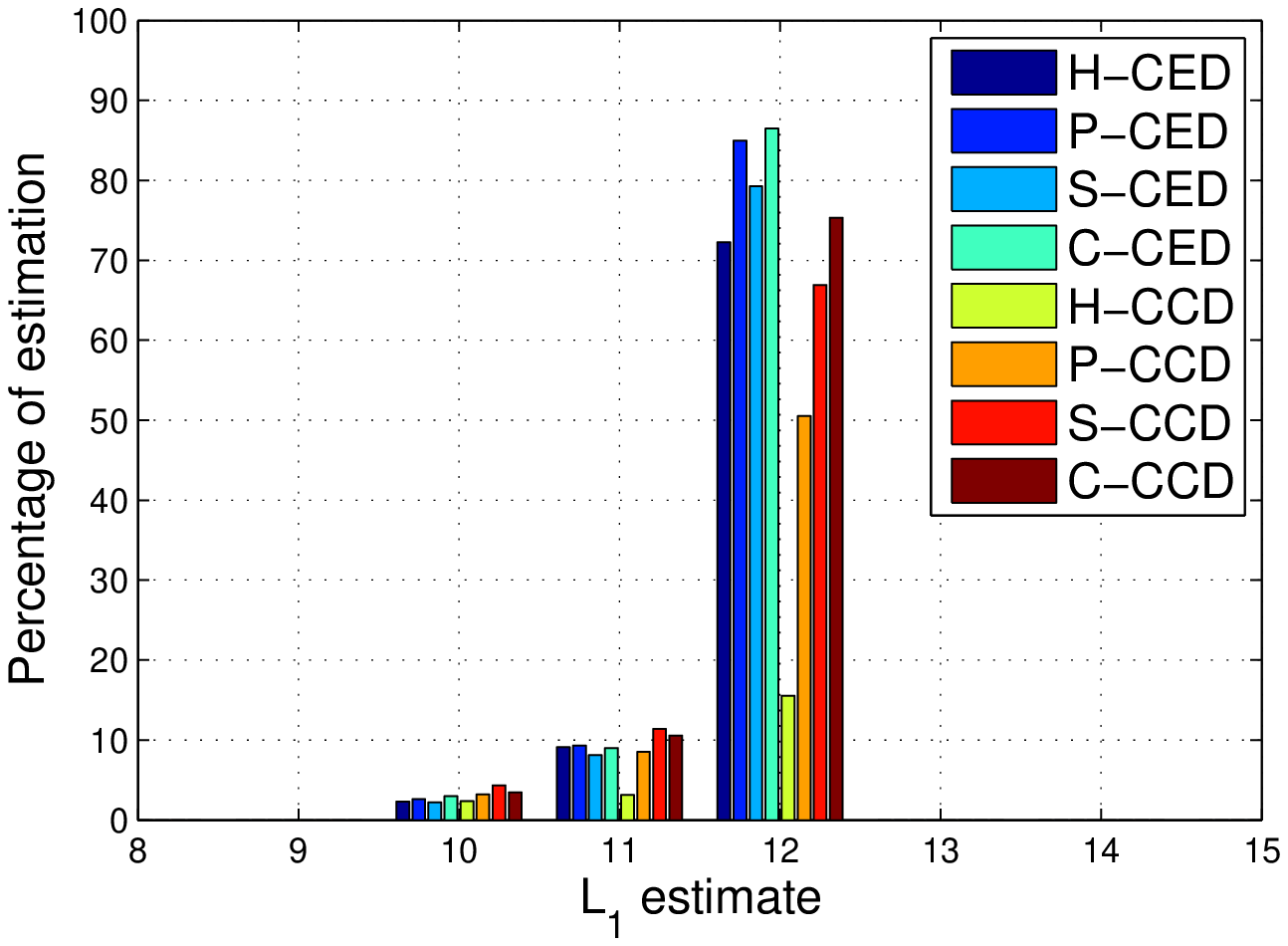}}
	\caption{Percentage of estimation of $L_1$ for the considered architectures assuming $N=6$ and a window of size $L=16$ starting from bin 4.}
	\label{fig_L1RealData}
\end{figure}

In Fig. \ref{fig_L1RealData}, we plot the percentage of estimation of $L_1$ based on the range bins 4-19. The histograms point out that the proposed architectures return $\widehat{L}_1=12$ with a greater percentage than their corresponding competitors.

\section{Conclusion}
\label{Sec:Conclusion}

In this letter, we have considered the problem of clutter edge detection and localization in training data. At the design stage, based on the GLRT, four architectures have been devised exploiting four possible CCM structures. 
The a priori knowledge of the CCM structure allowed us to extract more useful information from training data. Illustrative examples have shown the superiority of the proposed approaches over the competitors which ignore the CCM structure information. 
Future works might include the design of architectures that can classify the CCM structure or 
account for clutter discretes.

%


\IEEEtriggeratref{18}

\end{document}